\documentclass[aps,prmaterials,twocolumn,amsmath,amssymb,showpacs,superscriptaddress]{revtex4-2}
\usepackage{graphicx}
\usepackage{amssymb}
\usepackage{bm}
\usepackage{braket}
\usepackage{amsmath}
\usepackage{tabularx}
\usepackage{ltablex}
\usepackage{multirow}
\usepackage[dvipsnames]{xcolor}

\usepackage[colorlinks=true,
                linkcolor=blue,
	        citecolor=blue,
	        urlcolor=blue]{hyperref}

\newcommand{\trieste}{Dipartimento di Fisica, Universit\`a di Trieste, Strada Costiera 11, I-34151 Trieste, Italy}

\usepackage{verbatim}

\AtBeginDocument{\usepackage{booktabs}}
\begin{document}
\raggedbottom
\title{Thermal robustness of the quantum spin Hall phase in monolayer WTe$_2$}
\date{\today}
\author{Antimo Marrazzo}
\email{antimo.marrazzo@units.it}
\affiliation{\trieste}

\begin{abstract}
Monolayer 1T'-WTe$_2$ has been the first two-dimensional crystal where a quantum spin Hall phase was experimentally observed. In addition, recent experiments and theoretical modeling reported the presence of a robust excitonic insulating phase. 
While first-principles calculations with hybrid functionals and several measurements at low temperatures suggest the presence of a band gap of the order of 50 meV, experiments could confirm the presence of the helical edge states only up to 100 K. Here, we study with first-principle simulations the temperature effects on the electronic structure of monolayer 1T'-WTe$_2$ and consider the contributions of both thermal expansion and electron-phonon coupling. First, we show that thermal expansion is weak but tends to increase the indirect band gap. Then, we calculate the effect of electron-phonon coupling on the band structure with non-perturbative methods and observe a small reduction of the band inversion with increasing temperature. Notably, the topological phase and the presence of a finite gap are found to be particularly robust to thermal effects up to and above room temperature.
\end{abstract}

\maketitle
In 2014~\cite{xiaofeng_science_2014}, Qian et al.~predicted through first-principles simulations that two-dimensional (2D) transition metal dichalcogenides (TMDs) in the 1T' phase would exhibit a sharp quantum spin Hall effect (QSHE), characterized by strong band inversions and relatively large band gaps. These TMDs are defined by the chemical formula MX$_2$, where M is the transition metal (W or Mo) and X is the chalcogenide (Te, Se or S), and they exhibit a variety of polytypic structures such as 1H, 1T, and 1T'. 
In particular, the 1T structure in MX$_2$ TMDs is typically unstable in freestanding conditions and undergoes a spontaneous lattice distortion in the $x$ direction to form a period-doubling $2\times1$ distorted superstructure, the 1T' structure, made of 1D zigzag chains along the $y$ direction~\cite{calandra_prb_2013}.
In all the TMDs, but WTe$_2$~\cite{wilson_tprime_1969}, the 1T' structure is dynamically stable but it does not correspond to the lowest energy polymorph~\cite{xiaofeng_science_2014} and it is a metastable phase, although it can be stabilized under appropriate conditions~\cite{calandra_prb_2013} thanks to the large energy barrier between the 1H and the 1T' phases~\cite{xiaofeng_science_2014}. In WTe$_2$ instead, the 1T' structure corresponds to the most stable phase~\cite{wilson_tprime_1969,calandra_prb_2013}. Conductance experiments have confirmed the presence of a quantum spin Hall insulating (QSHI) state in 1T'-WTe$_2$ until 100 K~\cite{fei_wte2_natphy_2017,cava_science_2018}.
\begin{figure*}[!ht]
  \centering
  \includegraphics[width=0.33\linewidth]{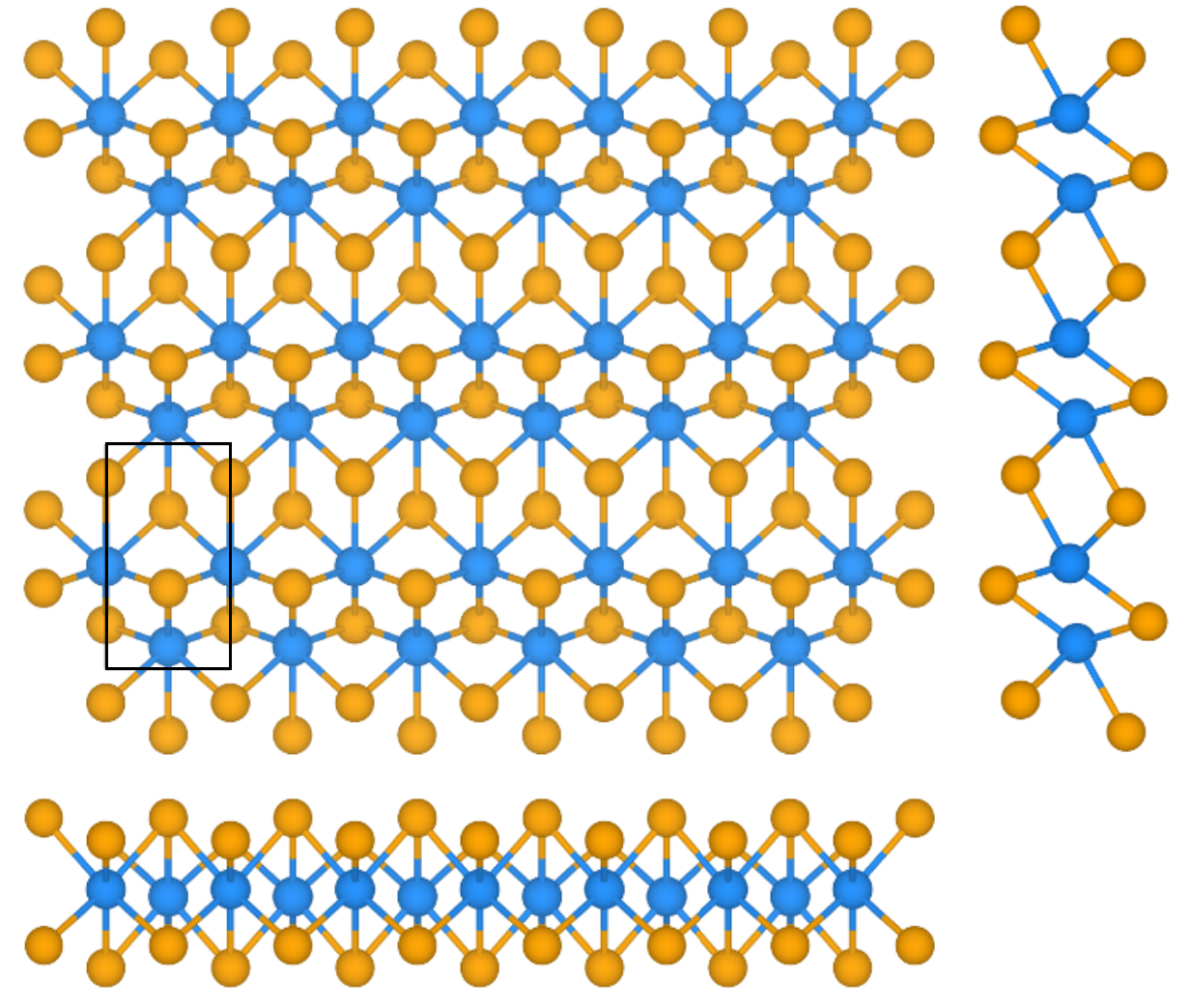}
  \put(-180,140){a$)$}
  \includegraphics[width=0.45\linewidth]{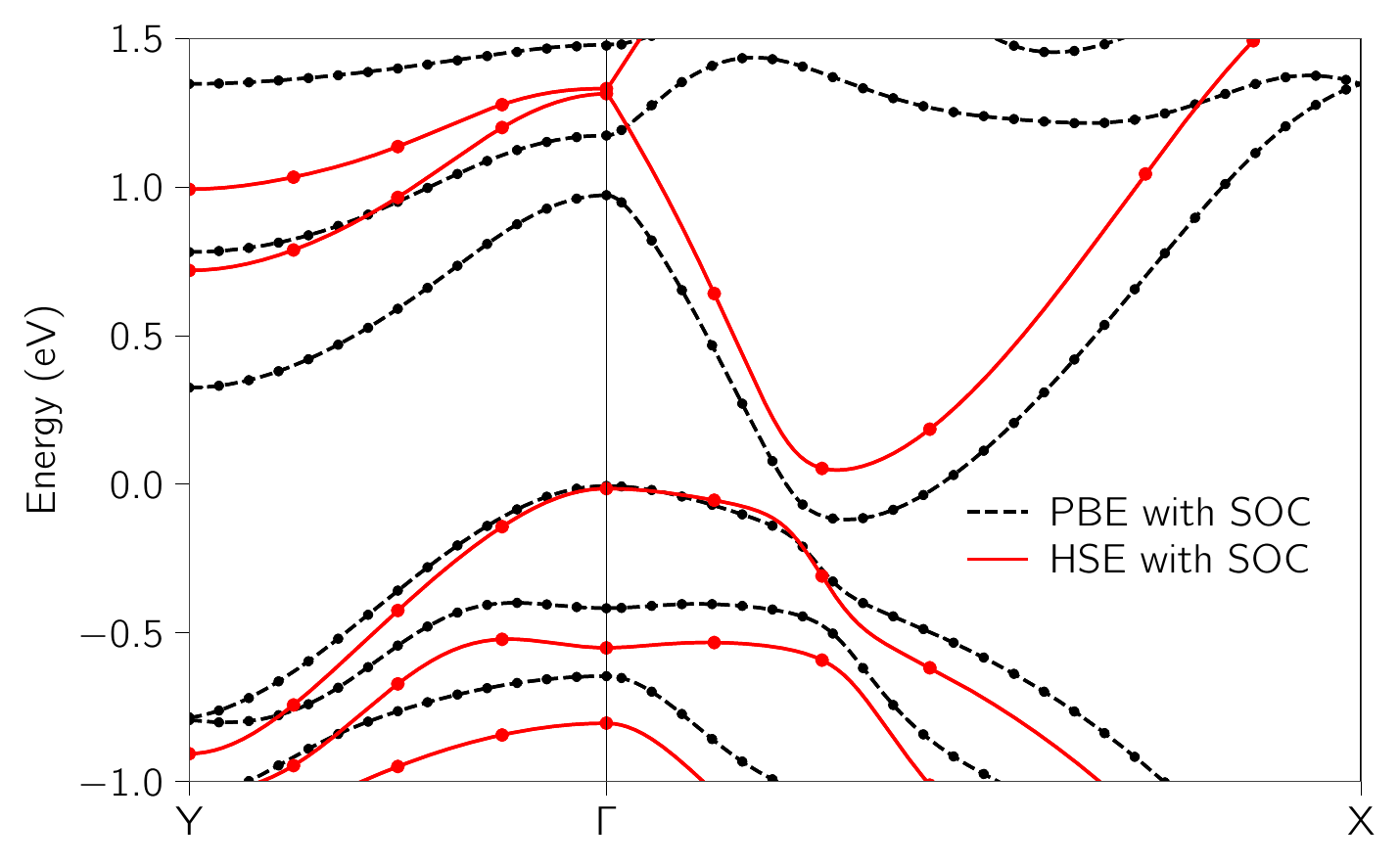}
  \put(-220,140){b$)$}
  \caption{Left panel (a): top and lateral views of monolayer 1T'-WTe$_2$, tungsten (tellurium) atoms are marked in light blue (orange) and the primitive call is drawn. Right panel (b): band structure for monolayer 1T'-WTe$_2$ with spin-orbit coupling (SOC) calculated at the DFT-PBE (black) and HSE (red) level. The energy zero is at top of the valence band, circles denote direct calculations while solid lines represent Wannier-interpolated states.}
  \label{fig:intro}
\end{figure*}

As QSHI material, 1T'-WTe$_2$ excels in many aspects. First, it exhibits a strong band inversion~\cite{xiaofeng_science_2014} of about 1 eV and a band gap above $k_BT_{R}$~\cite{cucchi_arpes_nanolett_2019},  where $T_{R}$ is room temperature. In addition, WTe$_2$ is a simple binary compound, which simplifies the experimental synthesis--especially with bottom-up approaches--compared to the numerous ternary QSHI compounds~\cite{marrazzo_nanolett_2019}. Finally, WTe$_2$ has a layered crystal structure and the binding energy between the layers is relatively low~\cite{mounet_two-dimensional_2018} (around 30 meV$/$\AA$^2$), such that the material is exfoliable into monolayers~\cite{morpurgo_natcom_2015}. It is then compelling that monolayers of 1T'-WTe$_2$ have become among the most promising 2D materials to realize the QSHE and the target of several experimental investigations~\cite{fei_wte2_natphy_2017,tang_wte2_natphys_2017,cava_science_2018,zhao_expstrain_2020,jia_evidence_2022,sun_evidence_2022}.

Although predictions report an extremely robust band inversion and indirect band gap above room temperature, experiments could find signatures of the QSHI phase until 100 K only, suggesting a possible role of temperature on the band structure. Here we show that the topological phase and the presence of a finite gap are particularly robust to thermal effects up to high temperatures, suggesting that extrinsic effects or thermally activated bulk conductance~\cite{wte2_conductance_arxiv_2022} might be responsible for the measured transition temperature.

The presence of a finite indirect band gap in monolayer WTe$_2$ has been debated. Density-functional theory (DFT) simulations~\cite{xiaofeng_science_2014,tang_wte2_natphys_2017} with the semilocal PBE functional predict a metallic state, with bands overlapping at the Fermi level and a direct gap throughout the Brillouin zone (BZ). On the contrary, calculations~\cite{zheng_wte2_advmat_2016,cucchi_arpes_nanolett_2019,custodial_prb_2019} with the HSE hybrid functional predict an indirect gap of about 50 meV. The predicted band gap is also strongly sensitive to the lattice constant~\cite{pulkin_wte2gaplatconst_2017,cucchi_arpes_nanolett_2019,zhao_expstrain_2020}, where even a small amount of strain can open a gap as predicted at the DFT level with different functionals such as LDA, PBE and PBEsol~\cite{pulkin_wte2gaplatconst_2017} and confirmed experimentally~\cite{zhao_expstrain_2020}. In this work, we calculate the PBE and HSE band structure both with spin-orbit coupling (SOC) on the equilibrium structure relaxed at the DFT-PBE level, as reported in Fig.~\ref{fig:intro}. The Wannier-interpolated HSE band structure has a finite indirect gap of 62 meV, while at the PBE level the system is metallic.

Angle-resolved photoemission spectroscopy~\cite{cucchi_arpes_nanolett_2019} (ARPES) and scanning-tunneling microscopy and spectroscopy~\cite{tang_wte2_natphys_2017} (STM/STS) experiments indicate the presence of a finite gap of around 50 meV, in agreement with the hybrid-functional predictions. Some recent STM/STS experiments instead suggested the presence of a metallic band structure with a Coulomb gap induced by disorder~\cite{song_observation_2018}. Finally, more recent transport~\cite{sun_evidence_2022} and STM/STS~\cite{jia_evidence_2022} experiments, backed-up by many-body calculations~\cite{sun_evidence_2022,excitonic_pnas_2021,varsano_monolayer_2020}, provided strong evidence of the presence of an excitonic insulating state. In addition, Ref.~\cite{jia_evidence_2022} explicitly ruled out the scenario of a band insulator, even in the presence of strain.

The band structure of topological insulators is characterized by two distinct quantities, the indirect band gap and the band inversion~\cite{bernevig_book_2013,vanderbilt_book_2018}. The presence of a finite gap guarantees the insulating nature of the bulk and ensures that, in an undoped finite crystallite, electronic transport can occur only through the topologically-protected metallic edge states. In all QSHI, the band gap is entirely driven by SOC and it is typically very weak, on the order of tens to hundreds of meV~\cite{marrazzo_nanolett_2019}. However, the strength of the topological phase is more connected to the size of the band inversion, which also determines the localization of the edge states~\cite{glide_prb_2019}. Following Refs.~\cite{tang_wte2_natphys_2017,chloe_prb_2016}, we define the band inversion~\cite{bernevig_book_2013,vanderbilt_book_2018} for WTe$_2$ as the energy difference at $\Gamma$ (the high-symmetry point where the band inversion occurs) between the lowest unoccupied band d$_{yz}^+$, formed by a combination of tungsten d$_{yz}$ orbitals with positive overall parity, and the occupied d$_{z^2}^-$ state (with negative parity), the latter sits at about 0.5 eV below the valence band maximum (VBM). The $d-d$ band inversion in WTe$_2$ is driven by the 1T' distortion and it is present also in absence of SOC~\cite{tang_wte2_natphys_2017,chloe_prb_2016}.

Temperature is known to affect the band structure of semiconductors~\cite{giustino_review_2017}, through the combined effects of thermal expansion (TE) and electron-phonon coupling (EPC). Both effects can occur also at zero temperature owing to the presence of zero-point motion, that can influence both the equilibrium lattice constant and renormalize the band gap. The inclusion of these effects can lead to rather different predictions with respect to calculations performed on the average atomic configuration at zero temperature; for instance zero-point renormalization (ZPR) alone has been estimated to modify the band gap of diamond by 0.6 eV~\cite{antonius_prl_2014}. It is then compelling to assess how TE and EPC affect topological phases, and in particular the persistence of the small indirect gap. In fact, typical predictions assume negligible temperature effects and deduce critical temperatures based on the zero-temperature structure with \emph{averaged} atomic positions. However, temperature-dependent studies of the electronic structure of topological materials can be computationally rather challenging, especially because they involve the calculation of EPC by considering both SOC and, as in the case of WTe$_2$, beyond-DFT methods like hybrid functionals or many-body perturbation theory at the GW level. Hence, fully first-principles studies of temperature effects in topological insulators are scarce and limited to 3D structures~\cite{antonius_prl_2016,monserrat_prl_2016,temp-pbte_prm_2019,temp-zrte_prr_2019,temp-snte_prb_2020,temp-bitei_prr_2020,cheol_prb_2020}. Overall, the results of these recent studies~\cite{antonius_prl_2016,monserrat_prl_2016,temp-pbte_prm_2019,temp-zrte_prr_2019,temp-snte_prb_2020,temp-bitei_prr_2020,cheol_prb_2020} provide a general picture of temperature effects in 3D topological insulators, where there seems to be no general trend for the sign of the correction to the band gap: temperature can promote or suppress the topological phase, depending on the very details of the system.

As mentioned above, in WTe$_2$ the electronic bands around the Fermi level and, in particular, the presence of an indirect gap, are very sensitive to the lattice constant~\cite{pulkin_wte2gaplatconst_2017,cucchi_arpes_nanolett_2019,zhao_expstrain_2020}, suggesting a possible important effect of TE on the band structure at finite temperature. In absence of an applied external pressure, we can calculate the equilibrium structure at any temperature T by minimizing the Helmholtz free energy $F\left(\left\{a_{i}\right\},T\right) = U - TS$ with respect to all the independent geometrical degrees of freedom 
$\left\{a_{i}\right\}$. Monolayer WTe$_2$ is orthorhombic with a rectangular unit cell, hence we set $\left\{a_{i}\right\}=(a,b)$, where we name $a$ the shortest lattice parameter and $b$ the longest one, the latter corresponds to the period-doubling direction $y$.  We adopt the quasi-harmonic approximation (QHA), where the vibrational free energy F is written as in the case of a perfectly harmonic crystal but where anharmonic effects are included through the dependence of the phonon frequencies on the lattice parameters. The QHA free energy can be then calculated as

\begin{eqnarray}
\label{eq:F_QHA}
F\left(\left\{a_{i}\right\},T\right)&=& E\left(\left\{a_{i}\right\}\right) + F_{vib}\left(\omega_{\mathbf{q},j}\left\{a_{i}\right\},T\right) \\
&=& E\left(\left\{a_{i}\right\}\right)+ \sum_{\mathbf{q},j} \frac{\hbar \omega_{\mathbf{q},j}}{2} \nonumber \\
&& +  k_{B}T \sum_{\mathbf{q},j} \ln\left[1-e^{\frac{-\hbar \omega_{\mathbf{q},j}}{k_{B}T}} \nonumber \right]
\end{eqnarray}
where $E$ is the electronic ground-state energy, $F_{vib}$ is the vibrational energy and $\omega_{\mathbf{q},j}$ are the phonon frequencies. The sums run on all the wave vectors $\mathbf{q}$ in the Brillouin zone (BZ) and all the phonon branches $j$. The first sum is temperature independent and it describes the zero-point motion. 

\begin{figure}
   \includegraphics{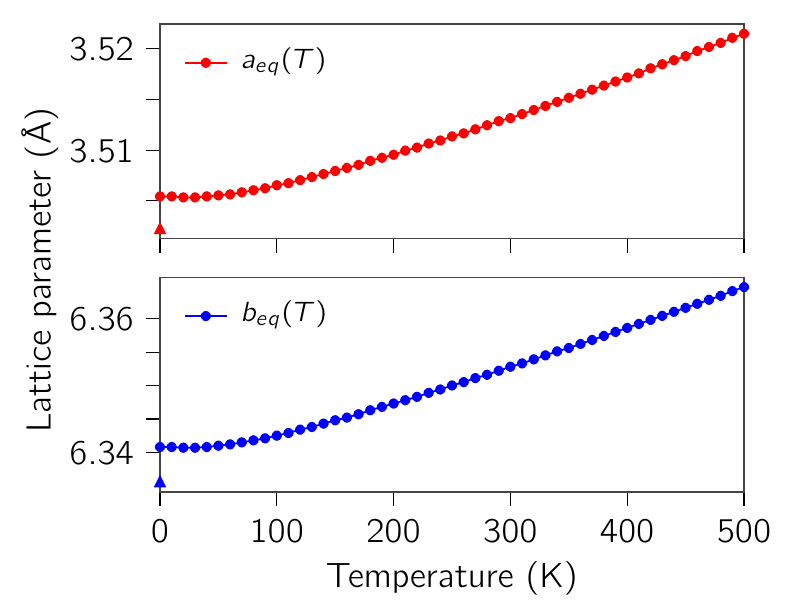}
   \caption{Equilibrium lattice parameters of monolayer WTe$_2$ from 0 to 500 K calculated with the quasi-harmonic approximation. The triangles at 0 K represent the equilibrium lattice parameters obtained without considering the zero-point motion.}
   \label{fig:qha}
 \end{figure}

For each temperature we calculate the minimum of the free energy on a dense grid of lattice constants and obtain the equilibrium lattice parameters $(a_{eq}(T),b_{eq}(T))$ (details in the SM~\cite{SM}). As experiments are performed at relatively low temperatures, we report in Fig.\ref{fig:qha} the results until 500 K, while we include the results until 1000 K in the SM~\cite{SM} for reference. In the temperature range that is more relevant for experiments, i.e. below room temperature, TE is very weak. The effect of ZPR is about $5\times 10^{-3}$\AA, while lattice constants increase by less than $0.1\%$ up to 150 K and of about $0.2\%$ at 300 K. The QHA predicts a $0.5\%$ ($0.4\%$) increase of the lattice constant $a$ ($b$) at 500 K.
The weak TE can be ascribed both to the small absolute change of the phonon frequencies with respect to the lattice constant and, especially at low temperatures, to the partial cancellation between negative and positive Gr{\"u}neisen parameters, although the effect is much weaker than in other 2D materials such as graphene for which a thermal contraction at low temperature has been predicted~\cite{mounet_prb_2005}.

Before discussing the results of the first-principles calculations and their interpretation, it is worth to discuss some more general considerations.
As already mentioned, the band gap of all QSHIs is always due to the presence of SOC, which gaps the band crossing. At finite temperature, it is tempting to assert that the system breaks  spatial symmetries at each point in time due to atomic vibrations, leading to a dynamical splitting of the degeneracies. However, in the harmonic approximation the symmetries are maintained upon thermal averaging as phonon-induced spin splitting are forbidden at equilibrium~\cite{dynrashba_prl_2021}. Only the presence of strong anharmonic effects, beyond the weak anharmonicity captured by the QHA, could break band degeneracies. Here we adopt the QHA and calculate the renormalization of the band structure due to the EPC by using the special displacement method (SDM)~\cite{prr_mario_2022}. The SDM is a non-perturbative approach in the adiabatic approximation, at each temperature the band structure is computed on a sufficiently large supercell where the atomic position are suitably distorted according to the phonon displacement patterns calculated with DFPT (details in the SM~\cite{SM}). 

\begin{figure*}
  \includegraphics[width=0.42\textwidth]{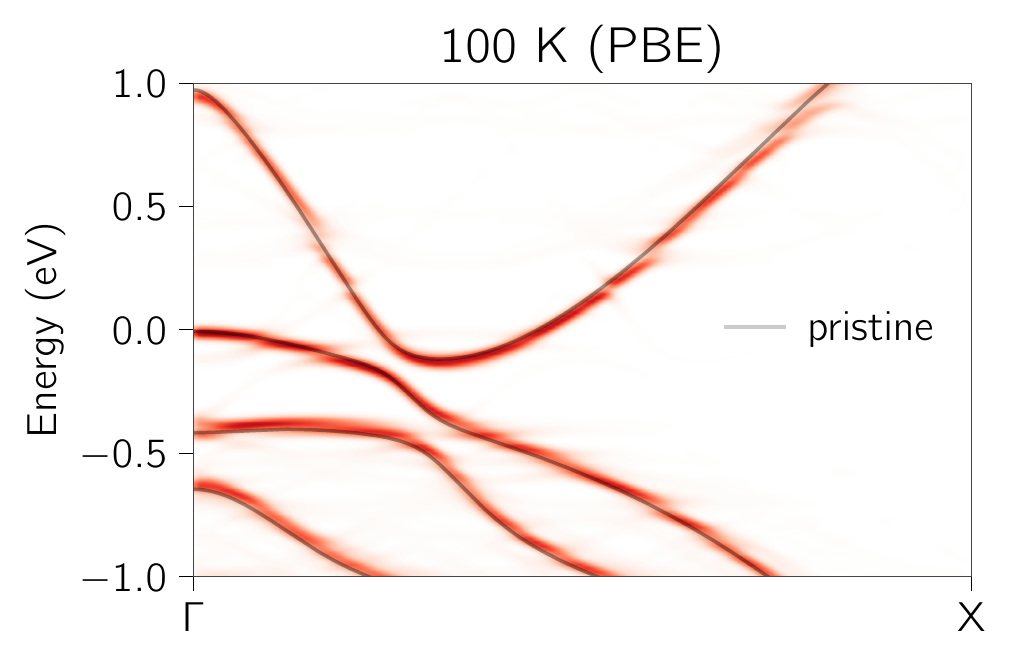}
  \put(-200,130){a$)$}
  \put(20,130){b$)$}
  \includegraphics[width=0.42\textwidth]{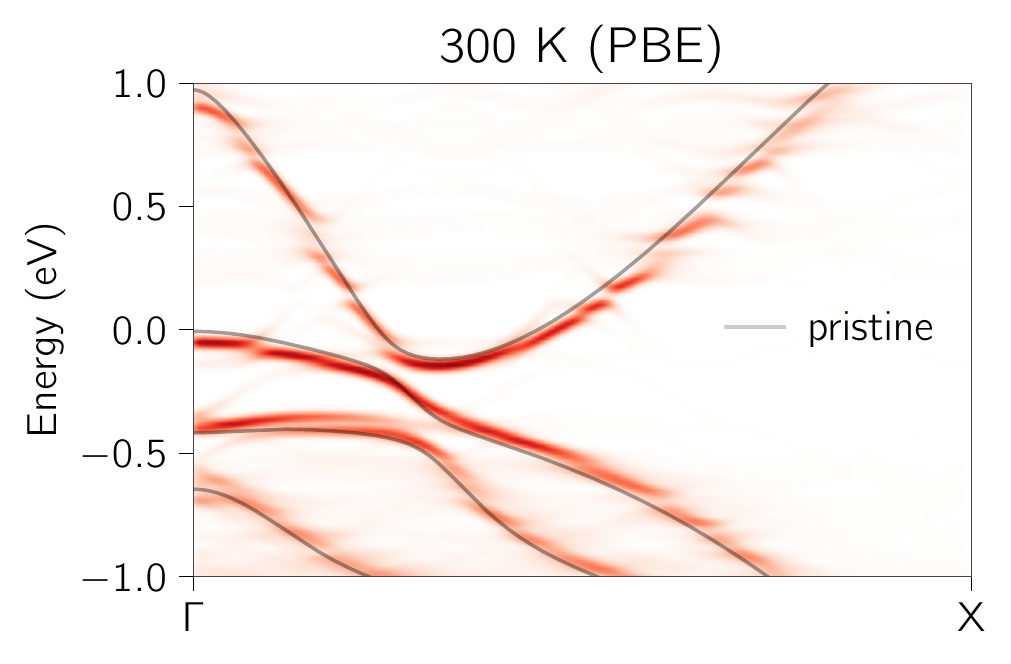}
  \includegraphics[width=0.42\textwidth]{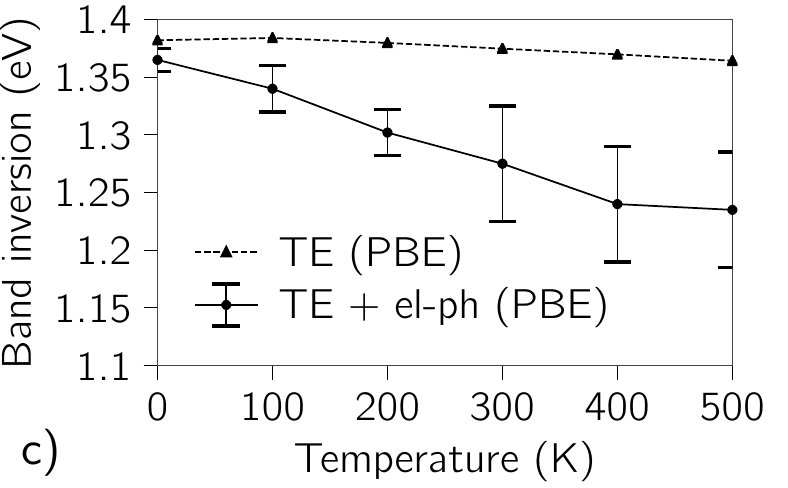}
  \put(-230,130){c$)$}
  \caption{Top panels: unfolded spectral density at a$)$ 100 K and b$)$ 300 K, the gray line marks the pristine calculation with atoms at the equilibrium positions. Bottom panel (c): temperature dependence of the band inversion. Calculations are performed with the special displacement method for a 108-atom supercell and the PBE functional with spin-orbit coupling. Triangles represent the contribution of thermal expansion (TE) only, while dots and error bars include also the effect of electron-phonon coupling (TE + el-ph).}
  \label{fig:spectral}
\end{figure*}

In Fig.~\ref{fig:spectral} we report finite-temperature band structures and band inversions at the PBE level obtained with the SDM and unfolded to the primitive BZ. The error bars account for the broadening of the band structure due to temperature and to the BZ unfolding, and include the spurious splitting of the degeneracies caused by the finite supercell. At 100 K, the band structure closely resembles the one obtained for the pristine monolayer where atoms sit at their equilibrium positions. In addition to thermal broadening, a weak reduction in the band inversion is present, while the finite direct gap is maintained. The effect becomes more visible at 300 K, where the reduction of the band inversion is more pronounced. We also report the strength of the band inversion from 0 K to 500 K (see Fig.~\ref{fig:spectral}), where we separate the contribution of TE from EPC. To sum up, TE has a negligible effect, while EPC leads to a reduction of the band inversion. The latter effect is relatively small and does not compromise the topological phase, but it seems sufficiently pronounced to be potentially measured with ARPES experiments.
Notably, our findings are compatible with ARPES measurements performed on 3D WTe$_2$ crystals, that reported that the band structure of bulk WTe$_2$ is temperature independent in between 20 and 130 K~\cite{temp-arpes-bulk-wte2_prb_2017}.

\begin{figure}
  \includegraphics[width=0.42\textwidth]{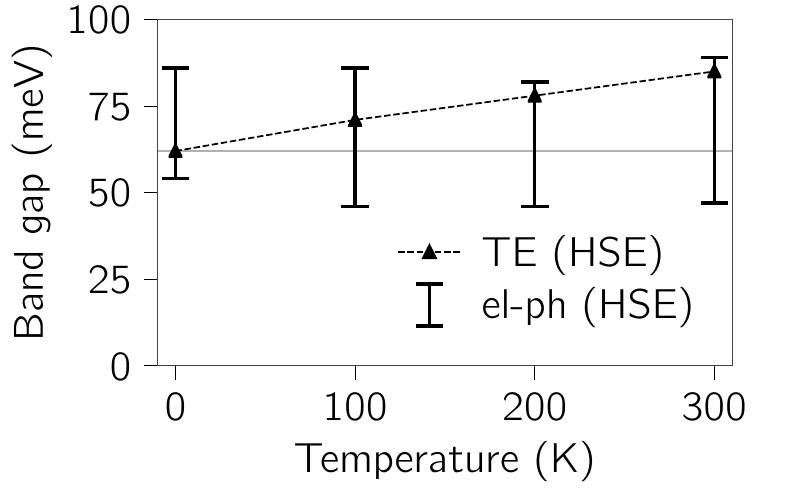} 
  \caption{Temperature dependence of the band gap. Calculations are performed with the special displacement method for a 108-atom supercell and the HSE functional with spin-orbit coupling. Triangles represent the contribution of thermal expansion (TE) only, while error bars include only the effect of electron-phonon coupling (el-ph). The grey horizontal line represents the reference HSE band gap calculated on the average atomic configuration at zero temperature.}
  \label{fig:hse_temp}
\end{figure}

While the strength of the topological phase of WTe$_2$ is due to a robust band inversion, the persistence of an insulating state with helical edge states is guaranteed as long as the band gap is finite. As semilocal functional, such as PBE, do not predict the presence of a finite indirect gap for the equilibrium structure, we deploy the SDM method with the HSE functional and SOC to study the temperature dependence of the band gap until 300 K, as reported in Fig.~\ref{fig:hse_temp}. Supercell calculations of 2D WTe$_2$ with hybrid functionals and SOC are computationally very intensive, and the magnitude of the band gap in WTe$_2$ is very small, of the order of tens of meV. Hence, we just aim to qualitatively probe the persistence of a gap as temperature is increased from zero to room temperature. In order to do so, we perform HSE calculations with SOC on 108-atom supercells, representing the uncertainties introduced by finite size effects with error bars (details in the SM~\cite{SM}). While for WTe$_2$ a meV-accurate quantitative estimate of the renormalized band gap seems to be beyond reach, the calculations clearly highlight the persistence of a indirect gap at high temperatures. In order to emphasize the robustness of the gap, we separately calculate the effect of TE and EPC. TE increases the band gap, going from 62 meV at zero temperature to 85 meV at 300 K. EPC does not seem to affect the indirect band gap, which never falls below 40 meV until 300 K, suggesting that the QSHI phase could in principle survive up to room temperature.

In Ref.~\cite{cava_science_2018} it was speculated that the 100 K transition temperature might not be an intrinsic limit, and that improvements in device quality may enable the observation of the QSHI phase at even higher temperatures. Notably, very recent theoretical work~\cite{wte2_conductance_arxiv_2022} suggests that actually moderate amounts of disorder can enhance the stability of conductance with respect to temperature, while the specific edge termination can affect the robustness of the topological protection.  Here, we have provided robust evidence that thermal effects on the electronic structure are not responsible for the disappearance of the topological phase with protected helical edge states. On the contrary, the QSHI is predicted to be very robust to the combined effect of TE and EPC, and a finite direct gap seems to persist until room temperature and possibly above. TE is very weak in free-standing WTe$_2$, although it leads to a small increase of the indirect band gap with temperature. EPC has a small but potentially measurable effect on the band inversion, the latter is reduced as temperature increases. Notably, TE seems to affect the indirect gap and not the band inversion, while the opposite is true for electron-phonon effects. The robustness of the band inversion strength implies that the remarkably small penetration length of the edge states is maintained at finite temperature.

Our findings rule out thermal effects within the harmonic and adiabatic approximation as possible explanations for the relatively low critical temperature at which pure edge conductance is suppressed in monolayer WTe$_2$. While the QSHI phase seems to persist at room temperature and above, it is possible that at about 100 K the conductance starts to be dominated by contributions of bulk states due to thermal broadening; this is the scenario suggested in Ref.~\cite{wte2_conductance_arxiv_2022} for ultra-clean samples.
Beyond bulk conductance and extrinsic effects, other intrinsic effects related to temperature remain to be investigated, as well as the role of the harmonic and adiabatic approximations. First, the TE of WTe$_2$ might be affected (e.g. the $a/b$ ratio) or even enhanced by the presence of a substrate. In addition, while in the harmonic approximation the band degeneracies due to crystalline symmetries are preserved at any temperature~\cite{dynrashba_prl_2021}, anharmonicity might affect the band structure~\cite{zach_prb_2020,zach_arxiv_2022} and lead to a physical splitting of the bands. On a different note, failures of the adiabatic approximation might lead to non-adiabatic EPC effects on the band structure, especially in the case of polar materials~\cite{miglio_npjcompmat_2020}. As WTe$_2$ is a centrosymmetric material made of heavy chemical elements, anharmonicity and non-adiabaticity are expected to be relatively weak; a quantitative computational study of their possible role will be the subject of future work. 

The author would like to thank Marco Gibertini for very useful discussions, and Feliciano Giustino and Mario Zacharias for clarifying crucial aspects of the special displacement method. The author acknowledges support from University of Trieste through the ``Microgrant 2021'' program and from CINECA, under the ISCRA initiative, for the availability of high-performance computing resources and support. Simulation time on MARCONI100 at CINECA was provided by the ISCRA-B project TEMPESTA, the ISCRA-C TEMPWTE2 and the CINECA-UniTS agreement.
\bibliography{biblio}
\section*{Supplementary Material}
\renewcommand{\figurename}{Supplementary Fig.}
\renewcommand{\tablename}{Supplementary Tab.}
\renewcommand\thefigure{\arabic{figure}}
\renewcommand\thetable{\arabic{table}}

\renewcommand\thesubsection{Supplementary Note \arabic{subsection}}
\subsection*{Methods}
\subsubsection*{\label{sm-fp}First-principles simulations}
Density-functional theory (DFT) calculations are performed with the Quantum ESPRESSO distribution~\cite{Giannozzi2017} using the PBE functional~\cite{perdew_pbe_96}. Structural relaxations are performed without spin-orbit coupling (SOC) using the SSSP Efficiency library and cutoffs v1.1~\cite{Prandini2018,gbrv}. The equilibrium structure without the effect of zero-point motion is obtained with variable-cell structural optimization and the BFGS algorithm. Band structures with SOC are obtained through fully-relativistic norm-conserving PseudoDojo~\cite{ONCVPSP,dojo_paper_18} pseudopotentials with an 80 (320) Ry wavefunction (charge density) cutoff. 
The Coulomb cutoff~\cite{Rozzi2006,Sohier2017} is used to avoid spurious interactions between periodic replicas and thus simulate the correct boundary conditions for 2D systems.  

Hybrid-functional calculations with SOC are performed using the Heyd-Scuseria-Ernzerhof (HSE) functional~\cite{HSE} and norm-conserving SG15~\cite{ONCVPSP,sg15} pseudopotentials, as implemented in Quantum ESPRESSO~\cite{Giannozzi2017} with the acceleration provided by the Adaptively Compressed Exchange Operator~\cite{lin_ace_2016}. HSE calculations for pristine WTe$_2$ are obtained with a 80 Ry cutoff for the wavefunction and the Fock operator, 320 Ry for the charge density, and with a k-grid of $14\times8\times1$ and a q-grid of $7\times4\times1$.

Band structures are interpolated using maximally-localized Wannier functions~\cite{wannier_review_2012} as implemented in Wannier90~\cite{Pizzi_2020}.

\subsubsection*{Phonons and the quasi-harmonic approximation}
\begin{figure}
    \includegraphics{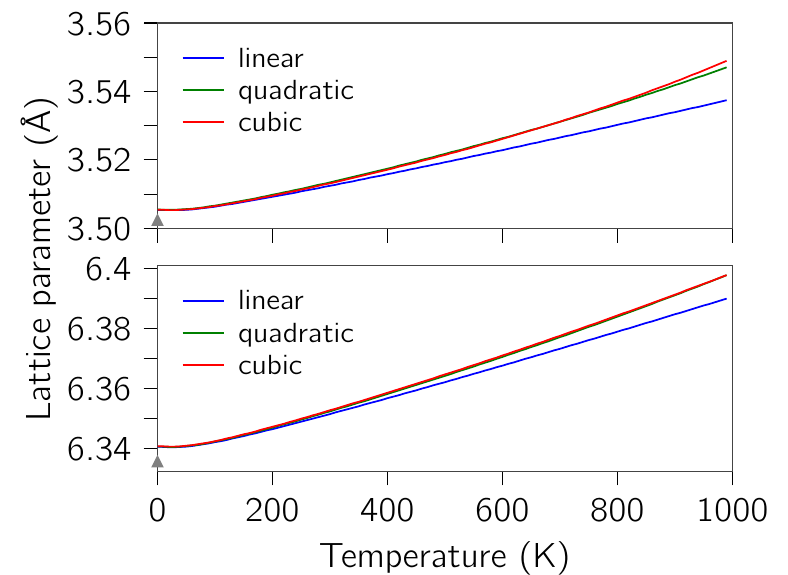}
    \caption{Equilibrium lattice parameters of monolayer WTe$_2$ from 0 to 1000 K calculated with the quasi-harmonic approximation. The three curves are obtained by considering a linear, quadratic and cubic polynomial for the fitting of the phonon frequencies with respect to the lattice constants. The triangles at 0 K represent the equilibrium lattice parameters obtained without considering the zero-point motion.}
    \label{fig:sm_qha}
  \end{figure}

Phonons are calculated on a 6$\times$4$\times$1 uniform q-point grid by using DFPT with pseudopotentials and cutoffs of the SSSP Efficiency library v1.1~\cite{Prandini2018,gbrv}. DFPT calculations are performed without SOC as the effect of the latter on the phonon frequencies is negligible; this is verified also by DFPT calculations at $\Gamma$ with and without SOC using the PseudoDojo pseudopotentials: the frequency difference due to SOC is never larger than $1.4\%$ on each mode individually. In the integration of the free energy, phonon frequencies are Fourier interpolated from the force constants to a 18$\times$12$\times$1 q-point grid. Phonons frequencies are calculated for 25 different lattice parameters that form a $5\times5$ uniform grid obtained by considering all the possible combinations of $a$ and $b$ strained by $0$, $\pm1\%$ and $\pm2\%$ around the zero-temperature relaxed configuration.~Then, the dependence of the phonon frequencies $\omega_{\mathbf{q},j}$ with respect to the $a$ and $b$ is fitted with a two-variable polynomial in order to calculate the vibrational energy $F_{vib}$ at any desired lattice constant. Linear, quadratic and cubic polynomials are considered, the results are reported in Fig.\ref{fig:sm_qha}. The cubic and quadratic polynomials give very similar results until 1000 K, while the linear fitting starts to deviate significantly only starting at 500 K. 

The electronic ground-state energy $E$ is fitted with a two-variable quartic polynomial and added to $F_{vib}$ obtain the Helmholtz free energy $F$. Finally, for each temperature the free energy is calculated on a dense grid in $(a,b)$ and minimized it to obtain equilibrium lattice parameters $(a_{eq}(T),b_{eq}(T))$. 

The equilibrium lattice constants represented by a triangle in Fig. 2 of the main text and in Supplementary Fig. 1 are obtained with variable-cell structural optimization and the BFGS algorithm.

\subsubsection*{Electron-phonon coupling and the special displacement method}
The renormalization of the band structure due to the electron-phonon coupling is calculated by using the special displacement method~\cite{prr_mario_2022} (SDM). The SDM is a non-perturbative approach where at each temperature the band structure is computed on a sufficiently large supercell constructed with atomic position which are suitably distorted according to the phonon displacement patterns calculated with DFPT. Thermal expansion is taken into account by calculating the phonons with the equilibrium lattice constants obtained in the QHA.
For improved accuracy, in the SDM we recalculate the phonons with a higher wavefunction and charge density cutoff, 80 Ry and 640 Ry respectively, using the SSSP pseudopotentials~\cite{Prandini2018,gbrv}. At each temperature, a 6$\times$3$\times$1 supercell of 108 atoms is constructed starting from the real-space force constants, by using the \textsf{ZG} code~\cite{prr_mario_2022} that is part of EPW~\cite{epw}. Supercell PBE and HSE calculations are performed with SOC and $\Gamma$ sampling. The wavefunction cutoff for supercell PBE calculations is equal to 80 Ry, while the corresponding cutoff for HSE calculations is set to 60 Ry and equal to the Fock cutoff.
In the study of the effect of electron-phonon coupling on the band inversion, we include error bars to represent the uncertainty in the determination of the inversion strength at the $\Gamma$ point. The bars account for the spectral broadening due to the unfolding procedure and include the intrinsic thermal broadening. Beyond the broadening, the spurious breaking of the band degeneracies is also included. The latter effect goes to zero in the infinite supercell limit, but it is always present in any tractable supercell. In the study of the indirect gap with HSE, we perform Wannier interpolation in the supercell BZ and represent with error bars the spurious splitting of the band degeneracies due to the finite supercell.

\end{document}